\nonstopmode
\documentstyle[12pt,world_sci]{article}
\def\beqn{\begin{eqnarray}{}}
\def\eeqn{\end{eqnarray}}
\def\oas#1{\hbox{${\cal O}(\alpha_s^{#1})$}}
\def\gsim{\lower0.5ex\hbox{$\stackrel{>}{\sim}$}}
\def\lsim{\lower0.5ex\hbox{$\stackrel{<}{\sim}$}}
\def\ff#1{\ifmmode{{}^{#1}F}\else{${}^{#1}F$}\fi}
\def\kk#1{\ifmmode{{}^{#1}{\rm K}}\else{${}^{#1}{\rm K}$}\fi}
\def\alphas{\alpha_s}

\def\fig#1{Figure~\ref{fig:#1}}
\def\mq{M_Q}

\input epsf.def
\epsfverbosetrue

\newtoks\test
\test{800}

\def\figi{
\begin{figure}[t]
 \epsfscale=800
  \centerline{\epsfbox{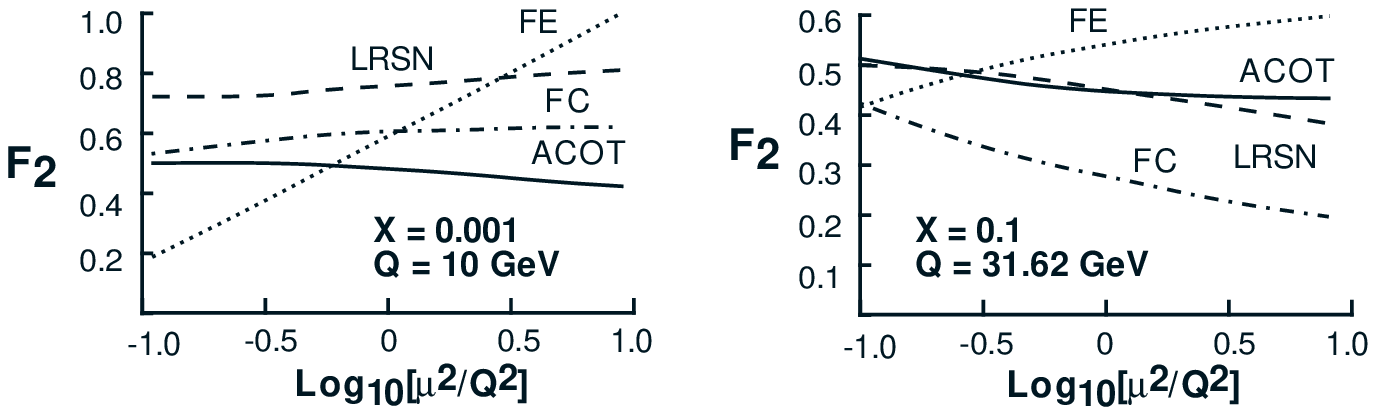}}
\vspace{-0.5cm}
      \caption{ \protect\small
The structure function $F_2(x,Q^2)$ vs. $\mu$ for charm production.
   }
   \label{fig:i}
\vspace{-0.5cm}
\end{figure}
}
\def\figii{
\begin{figure}[t]
 \epsfscale=800
  \centerline{\epsfbox{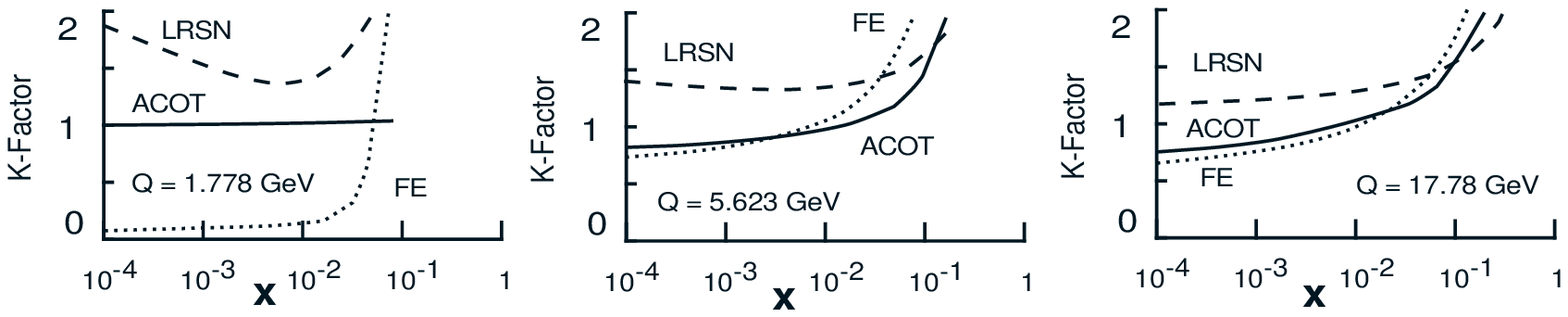}}
\vspace{-0.5cm}
      \caption{ \protect\small
The K-factor for the structure function $F_2(x,Q^2)$ vs. $x$ for
charm production.
   }
   \label{fig:ii}
\vspace{-0.5cm}
\end{figure}
}

\begin{document}
\null
\vfil

\begin{center}
\begin{tabular}{l}
August 1994 \\
\end{tabular}
    \hfill
\begin{tabular}{l}
SMU-HEP/94-22 \\
hep-ph/9409210
\end{tabular}
\\[1cm]

{\bf{\LARGE  Leptoproduction of Heavy Quarks \\
in the  Fixed and Variable Flavor
Schemes\footnote{Conference version of SMU-HEP/94-21}}}
\\[.5in]

 {\large
    Fredrick I. Olness\footnote{SSC Fellow}
   and Stephan T. Riemersma
  }
 \\[0.5in]
  Southern Methodist University,
   Dallas, Texas 75275
\end{center}
\vfil

\begin{abstract}

 We compare the results of the fixed-flavor scheme calculation of  Laenen,
Riemersma,  Smith and van Neerven with the variable-flavor scheme
calculation of Aivazis, Collins, Olness and Tung for neutral-current
(photon-mediated) heavy-flavor (charm and bottom) production.
 We compare the structure function $F_2(x,Q^2)$ throughout phase space,
and  also analyze the $\mu$-dependence
 We find that the former calculation is most applicable near
threshold, while the latter works well for asymptotic $Q^2$.

\end{abstract}

 \vfil

**Presented at the Eighth DPF Meeting, Albuquerque, NM, August, 1994**

\newpage

\title{{\bf  LEPTOPRODUCTION OF HEAVY QUARKS \\
IN THE  FIXED AND VARIABLE FLAVOR SCHEMES
}\footnote{Presented by F. Olness.
 Supported by  the  U.S. Department of Energy,
TNRLC,  and the Lightner-Sams Foundation. F.O. is supported in part by an SSC
Fellowship.
 The authors would like to thank J. Collins, J. Smith, D. Soper and W. Tung for
useful discussions.
}}
\author{FREDRICK I. OLNESS
   and STEPHAN T. RIEMERSM\\
{\em Southern Methodist University,
   Dallas, Texas 75275, USA}}

\maketitle
\setlength{\baselineskip}{2.6ex}

\begin{center}
\parbox{13.0cm}
{\begin{center} ABSTRACT \end{center}
{\small \hspace*{0.3cm}
 We compare the results of the fixed-flavor scheme calculation of  Laenen,
Riemersma,  Smith and van Neerven with the variable-flavor scheme
calculation of Aivazis, Collins, Olness and Tung for neutral-current
(photon-mediated) heavy-flavor (charm and bottom) production.
 We compare the structure function $F_2(x,Q^2)$ throughout phase space,
and  also analyze the $\mu$-dependence
 We find that the former calculation is most applicable near
threshold, while the latter works well for asymptotic $Q^2$.
}}
\end{center}

\section{Motivation of Variable and Fixed Flavor Schemes}

Several experimental groups
have studied the semi-inclusive deeply
inelastic scattering (DIS) process for heavy-quark production
$\ell_1(\ell_1) + N(P) \rightarrow \ell_2(\ell_2) + Q(p_1) + X(P_X)$.
Most analyses of this process assume that the hadron is comprised of only
the massless  gluon , up, down,  and strange  quarks, while
the charm, bottom, and top quarks are treated as massive
objects which are strictly external to the hadron.
 This view of the heavy quarks as external to the hadron
is appropriate when the energy scale of the process $\mu_{\rm phy}$
is not large compared to the mass of the heavy quark,
{\it i.e.} $M_Q \lsim \sqrt{s}$.
 With new data from HERA,
we can investigate the DIS process in a very different kinematic range from
that available at fixed-target experiments.  In this
new realm, the important question is:
Should the $c$ and $b$ quarks be considered as partons, or as heavy
objects extrinsic to the hadron?
Given that HERA extends the kinematic reach of the DIS process by
two orders of magnitude,
we can not expect our assumptions that were valid for fixed-target processes
to hold in a completely different kinematic regime.\cite{or}

Aivazis, Collins, Olness and Tung (ACOT) have discussed this issue at length
in reference~\cite{acot} and approach the problem by invoking
the {\it variable flavor scheme} (VFS), which varies the number of partons
according to the relevant energy scale $\mu_{\rm phy}$.
 The fundamental physical insight to the VFS is that in the region
$M_Q \,  \gg \, \mu_{\rm phy}$, the heavy quark should be {\it excluded}
as a constituent of the
hadron as it is kinematically inaccessible and decouples from the physics.
However, when
$M_Q \, \ll \, \mu_{\rm phy}$ the heavy quark should be {\it included} as a
parton since $\mq$ is insignificant compared to $\mu_{\rm phy}$.
Although the physics is unambiguous in these kinematic extremes,
most experimental data lies in between these clear-cut regions.
In the intermediate region, the renormalization scheme of Collins,
Wilczek and Zee (CWZ)
provides a well-defined transition between these two
extreme kinematic domains.

The above issue of what constitutes a parton also points to an
inconsistency  between traditional charged-current and neutral-current
heavy-quark production calculations~\cite{acot}.
When considering charged-current processes, one begins with  the purely
electroweak process $W + q \to Q$.
For neutral-current processes, the traditional approach is to begin
with  the \oas1 process $\gamma^* + g \to Q + \overline{Q}$
 As we work in the new kinematic regime spanned by HERA, the concept of
a ``heavy" quark becomes a relative term, and traditional distinction between
the charged-current and neutral-current calculations should vanish.
ACOT implements the CWZ renormalization and treats both charged-current
and neutral-current heavy-flavor production in a consistent fashion.

Laenen, Riemersma, Smith and van Neerven (LRSN) have calculated heavy-quark
production for DIS photon exchange, beginning with the \oas1 photon-gluon
fusion process and including the
complete \oas2 radiative corrections in reference~\cite{lrsn}.
  LRSN assumes there are no heavy-quark constituents to the hadron.
 For example, in producing $c$ quarks,
LRSN invokes only the  $g$, $u$, $d$ and $s$ partons.

\section{Comparison of Variable and Fixed Flavor Schemes}

\figii

In \fig{ii} we compare the $F_2$'s vs. $x$ at fixed $Q$.
 We refer to the curves as FE for the \oas0 flavor excitation  process,
FC for the \oas1 flavor creation process, ACOT for the complete VFS
calculation and LRSN for the complete FFS calculation.
 The FC curve is not shown as this is the denominator in the
definition of the K-factors. (Trivially, $K_{FC}=1$.)

For $Q = 1.778 \, GeV$,  the FE K-factor essentially vanishes as the
``heavy-quark" partons  should not contribute at low energy scales.
 The ACOT K-factor  approximately  reduces
to the FC result in the low energy limit.
 In this low $Q$-region, the LRSN calculation is the most
appropriate.

At $Q =  5.623  \, GeV$, the  very fast evolution of the FE
result makes it unreliable for  predicting  heavy-quark production.
 The ACOT K-factor now deviates from unity as the  fast evolution of the heavy
quark in the threshold region (due to  abundant gluons) generates important
contributions at relatively low values of $Q$.  However the subtraction
prescription ensures the result is reliable (in contrast to the FE process).
The LRSN K-factor decreases at small $x$, and flattens slightly.

At $Q =  17.78  \, GeV$, the FE, ACOT, and LRSN results have
similar shapes.   The K-factors are all  monotonically increasing vs. $x$.
In the range above $Q = 17.78 \,GeV$, the general characteristics are
similar.

In \fig{i} we compare the $F_2$'s vs. $\mu$ at fixed $\{x,Q\}$.
 For all values of $Q$, the FE process is increasing with  $\mu$ due to
the increasing heavy-quark PDF.  In contrast, the \oas1 FC process (driven by
gluons) is decreasing with $\mu$ largely due to the decrease in $\alphas(\mu)$.
The two-order calculations (ACOT and LRSN) that have compensating
contributions to cancel out some of the $\mu$-dependence.
Specifically, ACOT combines pieces of the FE and FC processes (together with a
subtraction term) to yield a result that has substantially less
$\mu$-dependence than either result in the large $Q$ region.  LRSN effectively
has the \oas2 FE contribution as the collinear heavy-quark part of phase
space is included, negating some of the $\mu$-dependence of the FC channel.

\figi

\section{Conclusions}

We have outlined the features of both the  VFS (ACOT) and
the FFS (LRSN) calculation. We summarize the highlights below.
   While the flavor excitation (FE) process can closely match the
two-order  results with a judicious choice of the scale $\mu$,
the large scale dependence  makes this unreliable.
    Likewise, while the flavor creation (FC) process is a good
starting point  in the threshold region.  However, the LRSN
calculation indicates that the  corrections to this naive
estimate can be  large.
   In the threshold region, the FFS (LRSN) calculation yields the
most stable and reliable results due to the domination of flavor
creation.
   In the asymptotic region, the VFS (ACOT)  calculation provides
the best results because of the dominance of the collinear
heavy-quark contribution.
 Furthermore, the ACOT result demonstrates that the
heavy-quark PDF's can yield  significant contributions at
relatively small scales,  ({\it i.e.} $\mu/\mq \sim 3$).

 We note that the difference between the LRSN and ACOT calculations above
threshold is suggestive of higher order contributions yet to be included.
 As such, the results of this comparison indicate that a combining of the LRSN
and ACOT calculations  in a consistent fashion (with  the additional mass
factorizations required) should allow us to make predictions based upon a
three-order result that combines the best attributes of both calculations.
 The result should be a calculation that will provide an important test of pQCD
when compared with the results from HERA.


\end{document}